\documentclass[11pt]{article}
\usepackage{epsfig}

\begin{document}
\setcounter{page}{1}

\begin{center}
\begin{Large}
{Local Polarimetry for Proton Beams with the STAR Beam Beam Counters}
\end{Large}

\vspace*{0.2in}
J. Kiryluk  for the STAR Collaboration \\
\vspace*{0.1cm}
{\footnotesize{\em{Massachusetts Institute of Technology \\ 77 Massachusetts Ave., Cambridge MA  02139-4307, USA\\E-mail: joanna@lns.mit.edu}}} \\

\end{center}

\vspace*{0.2in}

\begin{abstract}
STAR collected data in proton-proton collisions at $\sqrt{s}=200$ GeV with transverse and 
longitudinal beam polarizations during the 2003 running period at RHIC.
We present preliminary results on single transverse spin asymmetries $A_N$ 
in the production of forward charged hadrons detected with the Beam-Beam Counters (BBC).
The asymmetries $A_N$ measured at pseudorapidities $3.9 < |\eta| < 5.0$
are found to be of the order of $10^{-2}$ while 
asymmetries measured at smaller pseudorapidities $3.4 < |\eta| < 3.9$ 
are found to be consistent with zero.
The BBC and its associated scaler system for fast data recording provides the first local polarimeter at STAR.
The setup has been sucessfully used to tune the spin rotator magnets and to verify 
longitudinal polarization at STAR. 
\end{abstract}

\vspace*{0.2in}

STAR is one of two large experimental facilities at the Relativistic Heavy Ion Collider (RHIC) at 
Brookhaven National Laboratory.
One of the goals of the STAR physics program is to study the internal spin structure of the proton using 
collisions of longitudinally polarized protons at $\sqrt{s}=200-500$ GeV.
The stable beam polarization direction in RHIC is transverse, and thus needs to be rotated at 
the STAR Interaction Region (IR) to become longitudinal\cite{nim}.
The magnitude of the beam polarization is measured at a different IR by the RHIC Coulomb-Nuclear Interference 
(CNI) pC polarimeter\cite{nim}.
During the first (transversely) polarized proton running period in 2002
STAR has measured single transverse spin cross-section asymmetries $A_N=$
$(\sigma_{\downarrow}-\sigma_{\uparrow})/(\sigma_{\downarrow}+\sigma_{\uparrow})$ at $\sqrt{s}=200$ GeV in
neutral pion\cite{fpd} and charged hadron\cite{kiryluk} production in the forward region. 
Additional sets of magnets were installed before and after the STAR IR before 
the second proton running period at RHIC in 2003 to obtain longitudinal polarization.
Local polarimetry had been developed to measure the radial and vertical polarization components.
We discuss the instrumentation and results below. \\

\begin{figure}[ht]
\centerline{\epsfxsize=2.0in\epsfbox{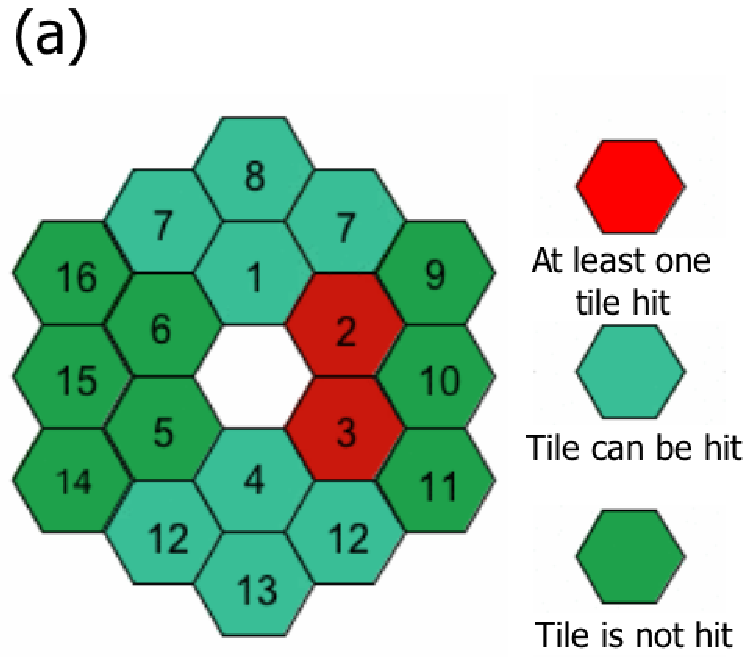}\epsfxsize=2.8in\epsfbox{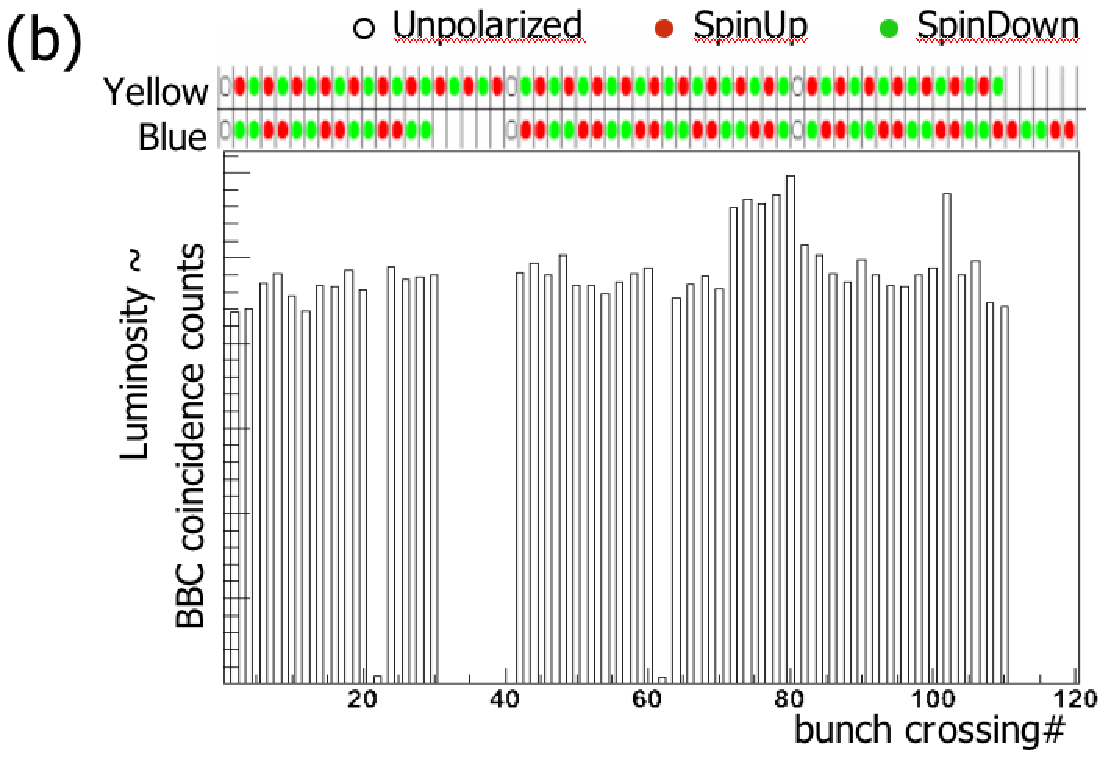}   }
\caption{ (a) Schematic view of the BBC small tiles and event topology for charged particle {\rm{Right}} scattering
at $3.9 < \eta < 5.0$. (b) The number of coincidences between the BBC on either side of the IR
versus bunch crossing for one data run. }
\label{fig:bbc}
\end{figure}

The Beam-Beam Counters (BBC) are scintillator annuli mounted around the beam pipe beyond the east and 
west poletips of the STAR magnet at 374\,cm from the nominal IR. 
The small tiles of the BBC shown in Fig.\ref{fig:bbc}(a) provide full azimuthal coverage $\Delta \phi = 2\pi$ 
in the pseudorapidity range of $3.4 < |\eta| < 5.0$. 
A coincident signal from any of the 18 tiles on the east side and any of the 18 tiles on the
west side of the IR constitutes a BBC coincidence.
The number of BBC coincidences is a measure of the luminosity $\mathcal{L}$.
The BBC acceptance amounts to 53\% of the total proton-proton cross section of $\sigma_{\rm{tot}}^{pp} = 51$ mb.
The BBC are used in STAR during proton runs to provide triggers, to monitor the overall luminosity, 
and to measure the relative luminosity for different proton spin orientations\cite{kiryluk}.
An example of the luminosity versus the bunch crossing number is shown in Fig.\ref{fig:bbc}(b). 
Each bunch crossing can be uniquely related to the relative spin orientations of the protons in the colliding beams.
The statistical accuracy of the relative luminosity measurement is typically $10^{-4}-10^{-3}$.
The high accuracy requires a fast BBC data acquisition system, consisting of a scaler board 
system\footnote{At average luminosities
$\mathcal{L}_{\rm{ave}} = 3.0 \times 10^{30} \rm{cm}^{-2} \rm{s}^{-1}$ the 
BBC coincidence rates are about~80~kHz,~which~are~orders~of~magnitude~higher than the STAR DAQ can handle.}.
The scaler board is a 24 bit and 10 MHz VME memory module\cite{hank} with
$2^{24}$ cells. Each cell is 40-bits deep to keep a continuous 
and deadtimeless record for up to 24 hours of operation at 10 MHz, which is the bunch crossing frequency at RHIC.
Seven of the 24 bits are used to keep the bunch crossing number.
The remaining 17 bits contain data from the fast detectors, such as the BBC.
The BBC data input to the scalers consist of discriminator outputs 
from 16 individual PMTs and logic levels produced by the STAR level 0 trigger electronics (a BBC coincidence signal).
In the spin asymmetry analysis described below, we used the data from two scaler boards for the BBC one 
each on the east and west side of the STAR IR.  \\

In an experiment with a transversely polarized beam and a left-right symmetric detector,
such as the BBC, the single spin asymmetry can be determined by measuring the beam polarization 
and the asymmetry of yields,
\begin{small}
\begin{equation}
\epsilon_{\rm{BBC}} = \frac{ r_{ij} - 1}{ 
r_{ij}+1 }
\simeq \left\{ 
\begin{array}{c}
A_N^{\rm{BBC}} \times P_{\rm{v}} \times \left<\cos{\phi}\right> \;\;  i,j={\rm{Left,Right}} \\
A_N^{\rm{BBC}} \times P_{\rm{r}}  \times \left<\sin{\phi}\right> \;\; i,j={\rm{Up,Down}}    \\
\end{array}
\right .
\label{eq:an}
\end{equation}
\end{small}
\noindent
in which $r_{ij}=\sqrt{ (N_i^{\uparrow} N_j^{\downarrow})/(N_i^{\downarrow} N_j^{\uparrow}) }$
and $N_{i(j)}^{\uparrow(\downarrow)}$ are the spin dependent yields
from the detector on the $i,j$=Left(Right) or Up(Down) side of the beam,
and $P_{v(r)}$ is the vertical (radial) beam polarization component. 
At RHIC both beams are polarized.
Therefore, one needs to sum the yields for both spin orientations of one beam
to measure a single spin asymmetry.
False asymmetries due to differences in luminosities and acceptences cancel 
in Eq.~\ref{eq:an}. 
The BBC segmentation allows the classification of the counted occurrences by pseudorapidity 
(2 bins: {\rm{Inner}} $3.9 < \eta < 5.0$ and {\rm{Outer}} $3.4 < \eta < 3.9$ BBC rings) and azimuth.  
The group of the 4 small tiles labeled by PMT numbers 1, 7 and 8 in Fig.~\ref{fig:bbc}(a) is referred to as {\rm{Up}},
whereas the group of tiles 4, 12, and 13 is called {\rm{Down}}.
The remaining small tiles are labeled {\rm{Left (Right) }} for the groups of tiles on the left (right).
An example of the hit topology for the {\rm{Inner-Right}} BBC event is shown in Fig.~\ref{fig:bbc}(a).

\begin{figure}[hb]
\centerline{\epsfxsize=3.3in\epsfbox{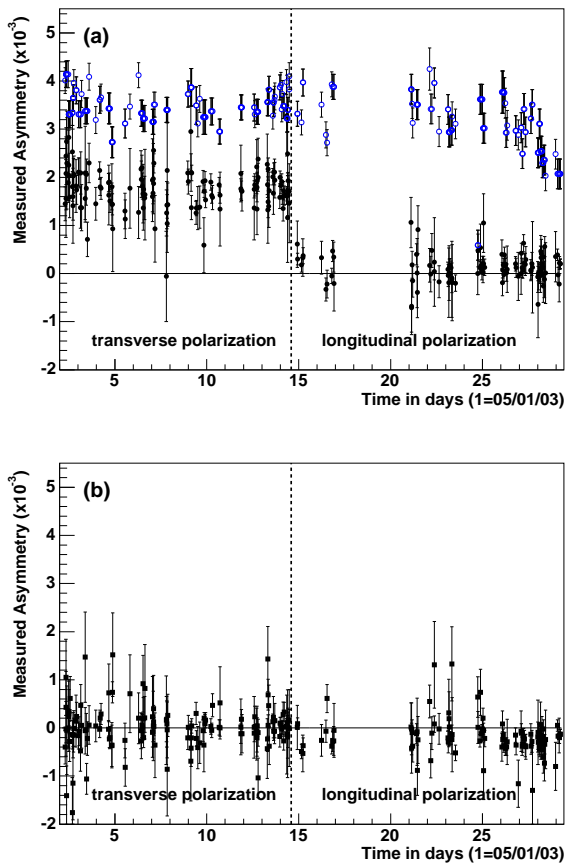}} 
\caption{ Measured (a) 'Left-Right' (b) 'Up-Down' asymmetries (in parts per thousand)
as a function of time (in days) since May 1 2003:
BBC asymmetries (filled points) and CNI asymmetries (open points).}
\label{fig:asym}
\end{figure}

Figure~\ref{fig:asym}(a) shows the time variation of the charged particle asymmetries determined with 
the BBC for $3.9 < \eta < 5.0$ (filled points) and the asymmetry measured with the RHIC CNI polarimeter\cite{cni} 
(open points). Each data point corresponds to one STAR run, which typically lasts for 30-60 min. 
The indicated uncertainties on the CNI and BBC asymmetries are statistical only.
The dashed line indicates when the spin rotators at STAR were turned on and the 
transverse polarization direction in RHIC was made longitudinal at the STAR IR. 
From the data with transverse beam polarization at STAR we find that 
$A_{\rm{N}}^{\rm{BBC}} = 0.67(8)\times A_{\rm{N}}^{\rm{CNI}} \sim 1 \% $ for $ 3.9 < \eta < 5.0 $,
while for smaller pseudorapidities, $ 3.4 < \eta < 3.9 $, the BBC asymmetries are found to be 
$A_{\rm{N}}^{\rm{BBC}} = 0.02(9)\times A_{\rm{N}}^{\rm{CNI}}$ 
consistent with zero. The systematic uncertainty on the measured BBC asymmetry was estimated to be 
smaller than statistical uncertainty, $\delta\epsilon_{\rm{sys}} = 5\times10^{-5}$. 
The {\rm{Left-Right}} asymmetries in the BBC are sensitive to the transverse polarization component, 
cf. Eq.~\ref{eq:an}. \\

\begin{figure}[ht]
\centerline{\epsfxsize=3.3in\epsfbox{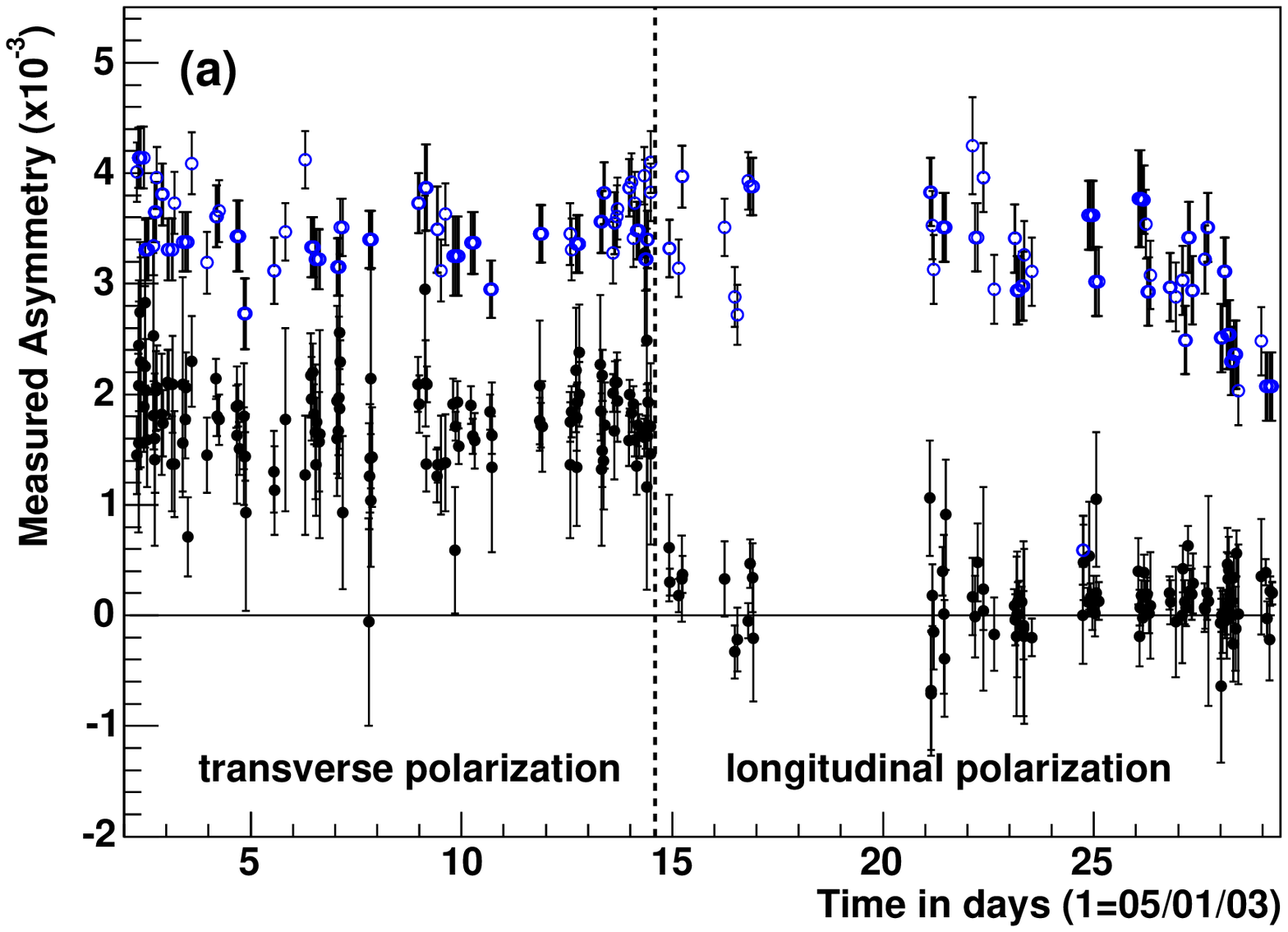}} 
\end{figure}
\vspace*{-0.8cm}

Their numerical values when the rotator magnets were on, that is, when the beam polarization was longitudinal 
at the STAR IR, were significantly smaller. The currents in the rotator magnets were adjusted to make these 
asymmetries consistent with zero, while at the same time the CNI polarimeter - located at a different RHIC IR 
- continued to measure non-zero beam polarization. The asymmetries have also been evaluated with the {\rm{Up}} 
and {\rm{Down}} groups of tiles in the BBC and are found to be close to zero for both transverse 
and longitudinal beam polarizations, cf. Fig.~\ref{fig:asym}(b), as expected.  \\

In summary, the BBC and its associated scaler system provided the first local polarimeter at STAR.
The measurement is based on the single transverse spin asymmetry $A_N$ in the detection of charged 
hadrons produced at forward rapidities, and is fast and non-destructive.
The setup has been sucessfully used to tune the spin rotator magnets at STAR and to monitor residual 
vertical and radial beam polarization components when the beam polarization was longitudinal. \\

\end{document}